\documentstyle[aaspp4,12pt,flushrt]{article}

\def\ifm#1{\relax\ifmmode#1\else$\mathsurround=0pt #1$\fi}

\newcommand{\apl}{\:^{<}_{\sim}\:}

\newcommand{\lya}{\mbox{${\rm Ly}\alpha$}}

\begin{document}

\title{\Large\bf A spectroscopically identified galaxy of probable redshift $z
= 6.68$}

\author{\large\bf Hsiao-Wen Chen\altaffilmark{1}, Kenneth M.
Lanzetta\altaffilmark{1}, \& Sebastian Pascarelle\altaffilmark{1}}

{\small \noindent $^1$Department of Physics and Astronomy, State University of
New York at Stony Brook, Stony Brook, NY 11794--3800, USA}

{\bf \noindent The detection and identification of distant galaxies is a
prominent goal of observational cosmology because distant galaxies are seen as
they were in the distant past and hence probe early galaxy formation, due to
the cosmologically significant light travel time.  We have sought to identify
distant galaxies in very deep spectroscopy by combining a new spectrum
extraction technique with photometric and spectroscopic analysis techniques.
Here we report the identification of a galaxy of redshift $z = 6.68$, which is
the most  distant object ever identified.  The spectrum of the galaxy is
characterized by an abrupt discontinuity at wavelength $\lambda \approx 9300$
\AA, which we interpret as the \lya\ decrement (produced by intervening
Hydrogen absorption), and by an emission line at wavelength $\lambda \approx
9334$, which we interpret as \lya.  The galaxy is relatively bright, and the
ultraviolet luminosity density contributed by the galaxy alone is almost ten
times the value measured at $z \approx 3$.}

  At near-infrared wavelengths, where background sky light is the dominant
source of noise, the Hubble Space Telescope (HST) using the Space Telescope
Imaging Spectrograph (STIS) is more sensitive than the Keck telescope because
(1) the sky is considerably fainter from space and (2) considerably higher
spatial resolution is attained from space, thus admitting far less
contaminating sky light.  To exploit the unique sensitivity of STIS at
near-infrared wavelengths, the Space Telescope Science Institute and the STIS
instrument team at the Goddard Space Flight Center initiated the STIS Parallel
Survey, in which deep STIS observations are obtained in parallel with other
observations$^1$.  We selected for analysis very deep observations acquired in
slitless spectroscopy mode (which records the dispersed light of an entire
field of view), because these observations are best suited for identifying
distant galaxies.

  The very deep observations were obtained by HST using STIS from 23 through 26
December, 1997 toward a region of sky flanking the Hubble Deep Field.  Most
observations consisted of a pair of images:  a direct image taken using no
filter and a dispersed image taken using the G750L grating.  Additional
observations consisted of only a direct image.  The integration time of the
direct images totaled 4.5 h over 82 exposures, and the integration time of the
dispersed images totaled 13.5 h over 60 exposures.

  We summed the direct and dispersed images using conventional image processing
techniques.  First, we reduced the images using standard pipeline software and
applied corrections for flat-field variations and illumination pattern.  Next,
we registered each pair of images to a common origin, using pointing offsets
determined from positions of bright stars in the direct images.  Next, we
measured the noise characteristics of the images, which are set by the
background sky level and the readout noise.  (The background sky level varies
by as much as a factor of two between individual exposures.)   Finally, we
summed the direct images to form a summed direct image and summed the dispersed
images to form a summed dispersed image, using optimal weights determined from
the measured noise characteristics and rejecting deviant pixel values caused by
cosmic ray events and hot pixels.  The spatial resolution of the summed direct
and dispersed images is ${\rm FWHM} \approx 0.08$ arcsec, the $1 \sigma$ single
pixel detection threshold of the summed direct image is $\approx 26.2$ mag
arcsec$^{-2}$, and the $1 \sigma$ single pixel detection threshold of the
summed dispersed image at $\lambda \approx 9800$ \AA\ is $\approx 5.5 \times
10^{-18}$ erg s$^{-1}$ cm$^{-2}$ \AA$^{-1}$ arcsec$^{-1}$. 

  We extracted one-dimensional spectra from the summed dispersed image using a
new spectrum extraction technique.  The extraction is made especially difficult
because the image of the field is covered by light of faint galaxies, which
when dispersed overlaps in a complicated way.  To overcome this difficulty, we
used the summed direct image to determine not only the exact locations but also
the exact two-dimensional spatial profiles of the spectra on the summed
dispersed image.  These spatial profiles are crucial because (1) they provide
the ``weights'' needed to optimally extract the spectra, and (2) they provide
the models needed to deblend the overlapping spectra and determine the
background sky level.  First,  we identified objects in the summed direct
image, using the SExtractor program of Bertin \& Arnouts$^2$.  Roughly 250
objects were identified in the summed direct image.  Next, we modeled each
pixel of the summed dispersed image as a linear sum of contributions from (1)
relevant portions of all overlapping neighboring objects and (2) background
sky.  The model parameters included roughly 250,000 ``object'' parameters (from
roughly 1000 spectral pixels each of roughly 250 objects) and 5000 ``sky''
parameters (from roughly 1000 fourth-order polynomials).  Finally, we minimized
$\chi^2$ between the model and the data with respect to the model parameters to
form one-dimensional spectra.  (In practice, we imposed a condition of
smoothness on the spectral variation of the sky in order to reduce the number
of sky parameters.  First, we smoothed the {\em model} of the sky by 10 pixels
in the spectral direction, and we subtracted this model from the data.  Next,
we performed the extraction again, but this time with sky subtraction turned
off.  In this way we reduced the effective number of sky parameters from
roughly 5000 to more like 500.  We found by experimentation that this number of
parameters is required to represent the spatial and spectral variations of the
sky.)

  We measured photometric redshifts from the one-dimensional spectra using a
variation of the photometric redshift technique described previously by
Lanzetta, Yahil, \& Fern\'andez-Soto$^{3,4}$ and Fern\'andez-Soto, Lanzetta,
\& Yahil$^5$.

  The results of the analysis are photometric redshift measurements of roughly
250 objects identified in the summed direct image.  Here we focus on one
particular galaxy identified by the analysis at a very high redshift, which we
designate as galaxy A.  The J2000 coordinates of galaxy A are $\alpha =
12:36:27.3$ and $\delta = +62:17:55.9$.  Interpretation of the spectrum of
galaxy A is especially straightforward because {\em this galaxy does not
overlap any other object in the dispersion direction}.  This means that in this
case much of the machinery of the extraction technique described above
(involving deblending overlapping spectra) does not actually come into play,
although the technique is required in order to accurately determine the
background sky level.

  Figure 1 shows the direct image of galaxy A and its neighbors in the top
panel.  The dispersed image of galaxy A and its neighbors and a map of the
detected emission lines are shown in the next two panels.  The dispersed image
of galaxy A with best-fit model spatial profiles of all objects but galaxy A
subtracted is shown in the next panel.  The one-dimensional spectrum of galaxy
A, a redshifted spectrum of a moderate-redshift ($z = 4.421$) galaxy$^6$, and
the one-dimensional spectrum of galaxy A cast into 325 \AA\ bins together with
the best-fit spectrophotometric template are shown in the next three panels.
Figure 2 shows the redshift likelihood function of galaxy A.  The spectrum of
galaxy A is characterized by (1) an abrupt discontinuity at wavelength $\lambda
\approx 9300$ \AA, with detectable continuum emission at wavelengths $\lambda =
9300 - 9950$ \AA\ and an absence of detectable continuum emission at
wavelengths $\lambda \apl 9300$ \AA, and (2) an emission line at wavelength
$\lambda = 9337 \pm 6$ \AA.  The redshift likelihood function of galaxy A is
characterized by a maximum at redshift $z = 6.84$ and a local maximum at
redshift $z = 6.66$, with the likelihood values of the primary and secondary
maxima statistically indistinguishable.  The energy flux of the emission line
is $2.6 \pm 0.5 \times 10^{-17}$ erg s$^{-1}$ cm$^{-2}$.  We interpret the
abrupt discontinuity as the \lya\ decrement and the emission line as \lya, in
which case the redshift of the galaxy determined from the emission line is $z =
6.68 \pm 0.005$.

  Several lines of evidence support the validity of the spectrum extraction and
the redshift identification:

  First, the integrated energy flux measured from the direct image is in
excellent agreement with the integrated energy flux measured from the dispersed
image.  The clear magnitude of galaxy A measured from the direct image is
$AB({\rm clear}) = 27.67 \pm 0.09$, and the clear magnitude of galaxy A measured
from the  dispersed image is $AB({\rm clear}) = 27.72 \pm 0.29$ (which is
obtained by integrating the product of the spectrum and the unfiltered
spectrograph and system throughput).  This demonstrates that the continuum
emission detected at wavelengths $\lambda \approx 9300 - 9950$ \AA---which is
crucial for establishing the redshift of galaxy A---is of exactly the amount
required to explain the direct image.  The clear magnitude of the emission line
alone of galaxy A measured from the dispersed image is $AB({\rm clear}) = 29.15
\pm 0.24$.  This demonstrates that the emission line alone {\em cannot} explain
the direct image; rather, additional continuum emission {\em must} be present,
namely the continuum emission detected at wavelengths $\lambda \approx 9300 -
9950$ \AA.  It is clearly unlikely that the analysis could have missed
significant continuum emission at wavelengths $\lambda < 9300$ \AA---where the
spectrograph and system throughput is relatively high---and instead detected
spurious emission (of exactly the amount required to explain the direct image)
at wavelengths $\lambda > 9300$ \AA.

  Next, various arguments demonstrate the reality of other faint emission lines
visible in the dispersed image.  For example, we applied the SExtractor program
to the smoothed dispersed image to objectively identify emission lines, setting
the detection threshold such that nothing was detected in the negative of the
image.  The resulting segmentation map is shown in the third panel of Figure 1.
Every emission line detected in the dispersed image by the SExtractor program
can be attributed to a galaxy detected in the direct image.  Specifically, the
bottom-most emission lines arise in a very faint galaxy, designated galaxy B in
Figure 1.  The spectrum of galaxy B is relatively uncomplicated, because the
galaxy just barely overlaps other galaxies above and below in the dispersion
direction.  Only very weak continuum emission is detected from the galaxy, but
at least three emission lines are clearly evident in the dispersed image:  an
emission line to the right and below the emission line of galaxy A, another
emission line in the same row just over 1/2 of the way from the left-hand edge
of Figure 1, and another emission line in the same row off the right-hand edge
of Figure 1.  The identifications of the three emission lines are (left to
right) Mg II $\lambda 2800$, He II $\lambda 3203$, and [O II] $\lambda 3727$ at
a redshift of $z = 1.213$, which is substantiated by comparison with the
ultraviolet spectrum of a starburst galaxy of Kinney et al.$^7$  Furthermore,
every emission line visible in Figure 1 (including another emission line below
and to the right of the emission line of galaxy A and a ``complex'' of emission
lines above and to the right of the emission line of galaxy A) is robust
against image processing and cosmic ray rejection methods and is exactly
coincident with the spatial profile of a galaxy detected in the direct image,
as is evident from the bottom image of Figure 1.

  Next, the abrupt discontinuity is consistent with interpretation as the \lya\
decrement but inconsistent with the interpretation as other spectral breaks
commonly observed in galaxy spectra.  The average energy flux density measured
from the dispersed image at wavelengths $\lambda = 7250 - 9250$ \AA\ is
$f_\nu(8250) = -0.01 \pm 0.04$ and at wavelengths $\lambda = 9400 - 9950$ \AA\
is $f_\nu(9675) = 0.67 \pm 0.22$ $\mu$Jy, which implies a decrement $1 -
f_\nu(8250) / f_\nu(9675) = 1.01 \pm 0.06$.  The corresponding $3 \sigma$ lower
limit to the break amplitude is 5.9, which exceeds by a significant factor the
largest measured break amplitudes $\approx 2.6$ of early-type galaxies and
$\approx 3$ of main-sequence stars (see Spinrad et al.$^8$).

  Next, the statistically insignificant but suggestive ``dip'' in the spectrum
of galaxy A at wavelengths between 9300 and 9500 \AA\ matches qualitatively the
expectation for stellar continuum radiation viewed through interstellar neutral
Hydrogen, which is expected to imprint a damped \lya\ absorption feature.
Specifically, the redshifted spectrum of the moderate-redshift galaxy shows a 
corresponding dip, as is evident from comparison of the middle and top 
one-dimensional spectra of Figure 1.  A similar feature is also present in the 
composite spectrum of 12 high-redshift ($z \approx 3$) galaxies presented by 
Lowenthal et al.$^9$

  Finally, the redshift determined from the emission line is in excellent
agreement with the redshift determined from the redshift likelihood function. 
Although difficult to quantify, the detection of an emission line at exactly
the wavelength predicted by the photometric redshift measurement lends {\em a
posteriori} support to the redshift identification.  In summary, the spectrum 
of galaxy A matches exactly the spectrum expected of a galaxy of redshift
approaching $z = 7$ but is unlike the spectra expected of lower-redshift
galaxies.

  The most striking property of galaxy A is that it is relatively bright at
wavelengths longward of \lya.  Although galaxy A is exceedingly faint in any
ordinary optical bandpass (due to very strong absorption by the \lya\ forest),
its continuum energy flux density at observed-frame wavelength $\lambda \approx
9800$ \AA\ or rest-frame wavelength $\lambda \approx 1300$ \AA\ is $f_\nu
\approx 1$ $\mu$Jy, which is comparable to the continuum energy flux densities
at similar rest-frame wavelengths of galaxies identified by Steidel and
collaborators$^{10}$ at redshifts $z \approx 3$.  This corresponds to an
unobscured star formation rate of $\approx 17 \ (147) \ h^{-2}$ $M_\odot$
yr$^{-1}$ for $q_0 = 0.5 \ (0.0)$, using the relationship between ultraviolet
luminosity and star formation rate with a Salpeter initial mass function of
Madau et al.$^{11}$.  Taking the redshift range searched for very high redshift
galaxies to extend from $z = 6$ to 7, the ultraviolet luminosity density
contributed by galaxy A alone is almost ten times the value measured at $z
\approx 3$.  If the galaxy is typical of the very high redshift galaxy
population, then apparently (1) very high redshift galaxies are luminous at
rest-frame ultraviolet wavelengths and (2) the unobscured cosmic star formation
rate may be substantially larger at $z \approx 7$ than at lower redshifts.

\vspace{0.1in}

\hrule

\begin{small}

\noindent Received {\underline{\makebox[1in]{}}.}

\noindent 1. Gardner et al.\ The STIS Parallel Survey: Introduction and First
Results.  Astrophys.\ J.\ {\bf 492}, L99--L102 (1998).

\noindent 2. Bertin, E. \& Arnouts, S. SExtractor: Software For Source
Extraction.  Astr.\ Astrophys.\ Suppl.\ {\bf 117}, 393--404 (1996).

\noindent 3. Lanzetta, K. M., Yahil, A., \& Fern\'andez-Soto, A. Star-forming
galaxies at very high redshifts. Nature {\bf 381}, 759--763 (1996).

\noindent 4. Lanzetta, K. M., Fern\'andez-Soto, A., \& Yahil, A. Photometric
Redshifts of Galaxies in the Hubble Deep Field in ``The Hubble Deep Field,
Proceedings of the Space Telescope Science Institute 1997 May Symposium,'' ed.\
M. Livio, S.  M. Fall, \& P. Madau (Cambridge:  Cambridge University Press) in
the press (1998).

\noindent 5. Fern\'andez-Soto, A., Lanzetta, K. M., \& Yahil, A. A new catalog
of photometric redshifts in the Hubble Deep Field. Astrophys.\ J., in the press
(1998).

\noindent 6. Steidel, C. C., Adelberger, K. L., Giavalisco, M., Dickinson, M.,
\& Pettini, M Lyman Break Galaxies at $z>4$ and the Evolution of the UV
Luminosity Density at High Redshift. Astrophys.\ J., submitted
(astro-ph/9811399).

\noindent 7. Kinney, A. L., Calzetti, D., Bohlin, R. C., McQuade, K.,
Storchi-Bergmann, T., \& Schmitt, H. R. Template Ultraviolet to Near-Infrared
Spectra of Star-Forming Galaxies and Their Application to K-Corrections.
Astrophys.\ J.\ {\bf 467}, 38--60 (1998).

\noindent 8. Spinrad, H. Stern, D., Bunker, A., Dey, A., Lanzetta, K., Yahil,
A., Pascarelle, S., \& Fern\'andez-Soto, A. A Sub--galactic Pair at $z = 5.35$.
Astr.\ J., {\bf 116}, 2617--2623 (1998).

\noindent 9. Lowenthal, J. D., Koo, D. C., Guzm\'an, R., Gallego, J., Phillips,
A. C., Faber, S. M., Vogt, N. P., Illingworth, G. D., \& Gronwall, C. Keck
Spectroscopy of Redshift $z$ Approximately 3 Galaxies in the Hubble Deep Field.
Astrophys.\ J.\ {\bf 481}, 673-688 (1997).

\noindent 10. Steidel, C., Giavalisco, M., Pettini, M., Dickinson, M., \&
Adelberger, K.  Spectroscopic Confirmation of a Population of Normal
Star-forming Galaxies at Redshifts $z > 3$. Astrophys.\ J.\ {\bf 462}, L17--L21
(1996).

\noindent 11. Madau, P., Pozzetti, L., \& Dickinson, M. The Star Formation
History of Field Galaxies. Astrophys.\ J.\ {\bf 498}, 106--116 (1998).

\end{small}

\acknowledgments

\noindent ACKNOWLEDGEMENTS.  We are grateful to H.\ Spinrad and to B.\ Woodgate,
B.\ Hill, and the rest of the STIS instrument team for important discussions. 
This research was supported by NASA and NSF.

\noindent CORRESPONDENCE should be addressed to H.-W.C. (email:
hchen@\-sbastr.\-ess.\-sunysb.\-edu).

\newpage

\figcaption{Image and spectra of galaxy A and its neighbors.  Top panel shows
the direct image of galaxy A and its neighbors, with arrows pointing to
galaxies A and B.  Angular extent of the image is 51 arcsec wide by 2 arcsec
high.  The second panel shows the dispersed image of galaxy A and its
neighbors, with arrows pointing to \lya\ emission line of galaxy A.  Pixel size
is 4.882 \AA\ in spectral direction and 0.05 arcsec in spatial direction,
resolution is roughly two pixels (in both directions), and spectrum is boxcar
smoothed by five pixels in spectral direction and two pixels in spatial
direction.  The third panel shows the map of the detected emission lines.
(Note that the continua of the topmost bright objects have been broken up by
the SExtractor program and are not individual emission lines.)  The forth panel
shows the dispersed image with best-fit model spatial profiles of all objects
but galaxy A subtracted.  The fifth panel shows the one-dimensional spectrum of
galaxy A.  Pixel size is 5 \AA, resolution is roughly two pixels, and spectrum
is boxcar smoothed by three pixels.  The sixth panel shows a redshifted (to $z
= 6.68$) spectrum of a moderate-redshift ($z =4.421$) galaxy.  Bottom panel
shows the one-dimensional spectrum of galaxy A cast into 325 \AA\ bins together
with best-fit spectrophotometric template spectrum.  Vertical error bars
indicate $1 \sigma$ uncertainties and horizontal error bars indicate bin sizes.
The images were recorded by a $1024 \times 1024$ CCD detector, which for the
direct images was operated in unbinned mode (resulting in a $1024 \times 1024$
grid) and for the dispersed images was operated in $1 \times 2$ binned mode
(resulting in a $1024 \times 512$ grid).  The spatial scale of the image on the
detector was 0.05 arcsec pixel$^{-1}$, which yielded a field of view of $51
\times 51$ arcsec$^2$.  The telescope was displaced by several arcsec between
each observation in order to reduce the effects of pixel-to-pixel sensitivity
variations.}

\figcaption{Redshift likelihood function of galaxy A.  We measured photometric
redshifts using a variation of the photometric redshift technique described
previously.  First, we adopted spectral templates of E/S0, Sbc, Scd, and Irr
galaxies, including the effects of intrinsic and intervening neutral hydrogen
absorption.  (These spectral templates, which are described by Lanzetta, Yahil,
\& Fern\'andez-Soto$^4$ and Fern\'andez-Soto, Lanzetta, \& Yahil$^5$, span
rest-frame wavelengths $\lambda = 912 - 22,500$ \AA.)  Next, we constructed the
``redshift likelihood functions'' by calculating the likelihood of obtaining a
measured spectrum given a modeled spectrum at an assumed redshift, maximizing
with respect to galaxy spectral type and arbitrary flux normalization.
Finally, we determined the maximum-likelihood photometric redshift measurements
by maximizing the redshift likelihood functions with respect to redshift.}

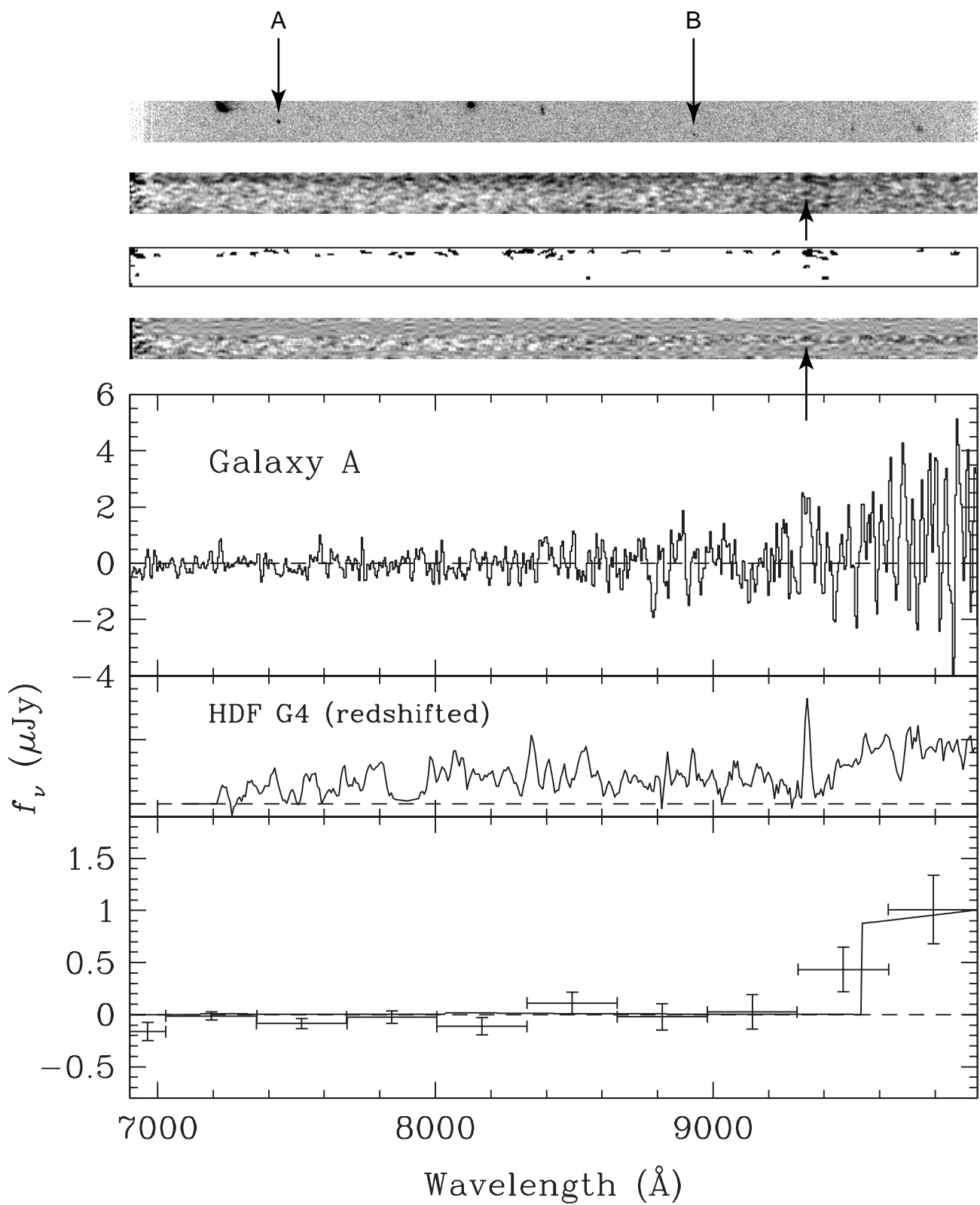
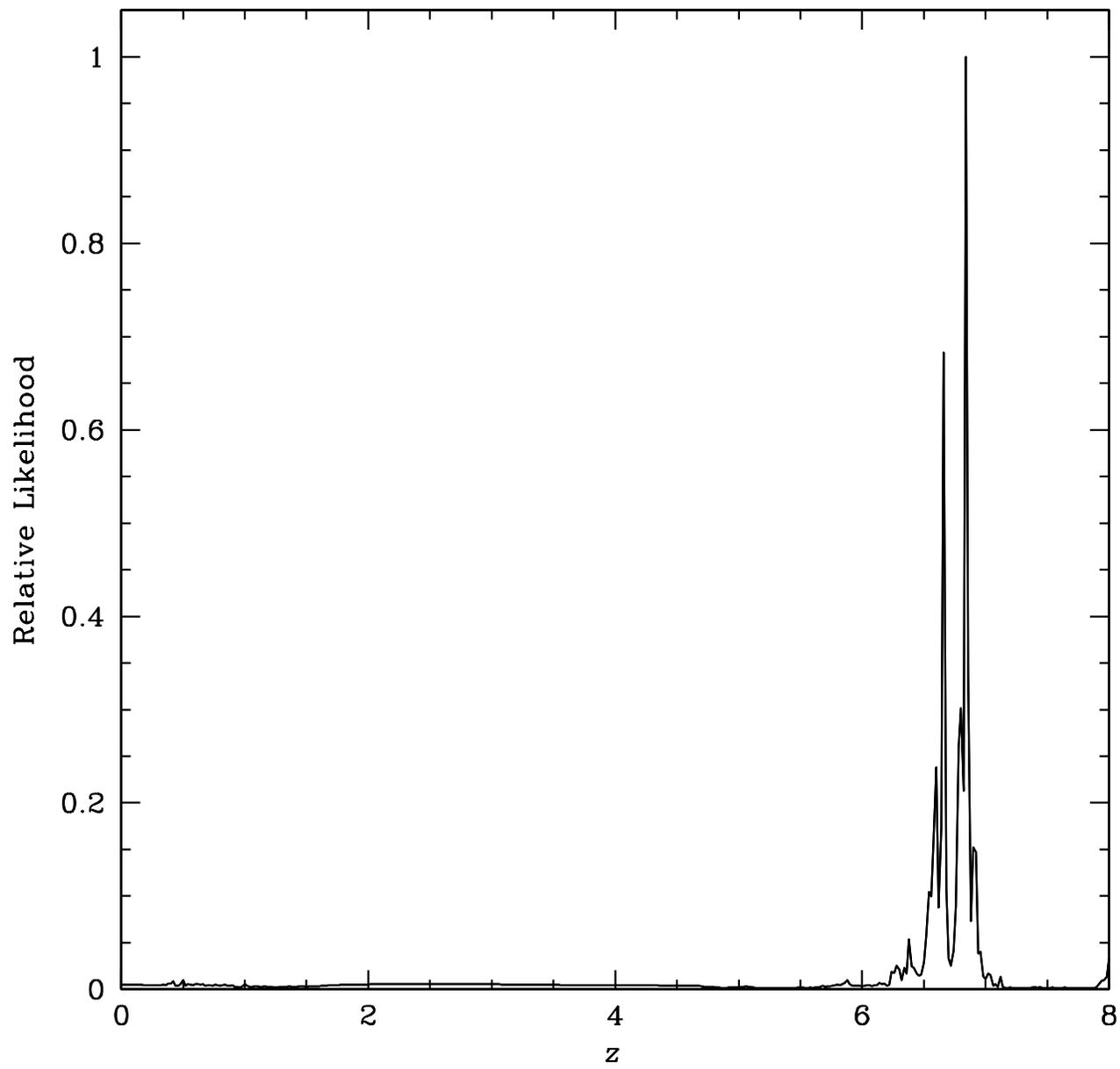
\end{document}